\documentclass[reprint,aps,pra,superscriptaddress]{revtex4-2}
\usepackage{bm,physics,amsmath,amssymb,
	graphicx,xcolor,embedfile,comment}
\usepackage[utf8]{inputenc}
\usepackage[normalem]{ulem}
\usepackage[T1]{fontenc}
\usepackage{tikz}

\newcommand{\UP}{\pmb{\Uparrow}}
\newcommand{\DOWN}{\pmb{\Downarrow}}

\begin{document}

\title{Effective model for studying optical properties of lead-halide perovskites}

\author{Artem G. Volosniev}
\affiliation{IST Austria (Institute of Science and Technology Austria),
    Am Campus 1, 3400 Klosterneuburg, Austria}
    
\author{Abhishek Shiva Kumar}
\affiliation{IST Austria (Institute of Science and Technology Austria),
    Am Campus 1, 3400 Klosterneuburg, Austria}

\author{Dusan Lorenc}
\affiliation{IST Austria (Institute of Science and Technology Austria),
    Am Campus 1, 3400 Klosterneuburg, Austria}
    
\author{Younes Ashourishokri}
\affiliation{IST Austria (Institute of Science and Technology Austria),
    Am Campus 1, 3400 Klosterneuburg, Austria}
    
\author{Ayan A. Zhumekenov}
\affiliation{KAUST Catalysis Center (KCC), Division of Physical Sciences and Engineering, King Abdullah University of Science and Technology (KAUST), Thuwal 23955-6900, Kingdom of Saudi Arabia} 
    
 \author{Osman M. Bakr}
\affiliation{KAUST Catalysis Center (KCC), Division of Physical Sciences and Engineering, King Abdullah University of Science and Technology (KAUST), Thuwal 23955-6900, Kingdom of Saudi Arabia}    
    
\author{Mikhail Lemeshko}
\affiliation{IST Austria (Institute of Science and Technology Austria),
    Am Campus 1, 3400 Klosterneuburg, Austria} 
    
\author{Zhanybek Alpichshev}
\email{alpishev@ist.ac.at}
\affiliation{IST Austria (Institute of Science and Technology Austria),
    Am Campus 1, 3400 Klosterneuburg, Austria} 
    
\begin{abstract}
We use
general symmetry-based arguments to construct an effective model suitable for studying optical properties of lead-halide perovskites. 
To build the model, we identify an atomic-level interaction between electromagnetic fields and the spin degree of freedom that should be added to a minimally-coupled $\mathbf{k\cdot p}$ Hamiltonian. 
As a first application, we study two basic optical characteristics of the material: the Verdet constant and the refractive index.  Beyond these linear characteristics of the material the model is suitable for calculating non-linear effects such as the third-order optical susceptibility. Analysis of this quantity shows that the geometrical properties of the spin-electric term imply isotropic optical response of the system, and that optical anisotropy of lead-halide perovskites is a manifestation of hopping of charge carriers. To illustrate this, we discuss third-harmonic generation.
\end{abstract}

\maketitle

Lead-halide perovskites (LHP) is a family of lead-based compounds with the structure APbX$_3$ where A=Cs, CH$_3$NH$_3$; X=Cl, Br, I. They attract attention as promising candidates for solar sells with high performance and stability~\cite{Gratzel2014,Saliba2016}. In order to construct improved photovoltaic devices as well as to explore further potential applications of perovskite compounds, it is necessary to gain a better understanding of the material properties of this system. The study of basic optoelectronic phenomena in LHP in the near-infrared range is particularly important in this regard as it may provide an insight into microscopic properties of the charged excitations in these materials such as dispersion and excitonic states~\cite{Green2015,Leguy2016}.

The response of a given system to applied electromagnetic fields crucially depends on how these fields couple to the (relevant) degrees of freedom of the system. In this paper, which accompanies~\cite{Volosniev2022}, we argue that introducing electromagnetic field by naive minimal electromagnetic coupling to the $\mathbf{k\cdot p}$ Hamiltonian does not adequately capture the frequency dependence of basic optical properties of lead-halide perovskites. Moreover, we amend the $\mathbf{k\cdot p}$ Hamiltonian by introducing new terms which have a transparent physical meaning. We test our effective description of LHP against available experimental data and produce numeric values for the strengths of the new terms in the effective Hamiltonian. 

The main results of the paper are as follows: First, we argue that the Hamiltonian introduced in Ref.~\cite{Volosniev2022} satisfies all necessary symmetries for describing low-energy optical properties of LHP. Second, we employ the model to calculate linear optical quantities. Finally, we argue that non-linear optical effects can also be conveniently studied within the introduced theoretical framework. We compute the third-order optical susceptibility, which (in particular) allows one to investigate polarisation dependent non-linear properties of LHP. As an illustration, we discuss anisotropy in third-harmonic generation (THG).   

The paper is structured as follows: in Sec.~\ref{sec:effective_description}, we use general symmetry-based arguments to construct an effective low-energy description of LHP. This result is then used in Secs.~\ref{sec:polarizability} and~\ref{sec:faraday} to calculate the linear optical polarizability and the Verdet constant. We show how to calculate non-linear effects using the effective model in Sec.~\ref{sec:non_linear}. Section~\ref{sec:summary} contains a brief summary of our work. Technical details to support the discussion are presented in three appendices.

 \section{Effective description of LHP}
 \label{sec:effective_description}
 
 \subsection{Basis states and Symmetries}
 
The low-energy optoelectronic properties of LHP are determined by the hybridization of $s$- and $p$-orbitals of Pb and halide atoms~\cite{Umebayashi2003}. Near the gap edge the states have a pronounced $s$- (valence band) and $p$-type (conduction band) character. As a result of spin-orbit coupling the $p$-type states further split into $J\!=\!1/2$ and $J\!=\!3/2$ manifolds. The former shapes the properties of the bottom of the conduction band. Therefore, to analyze the low-energy physics of LHP, it is sufficient to consider only the following four basis states~\cite{Becker2018}: 
\begin{align}
|\DOWN\Uparrow\rangle&=\left(|S\frac{1}{2},\frac{1}{2}\rangle\right) = |s\rangle |\uparrow\rangle \label{eq:states_1}\\
|\DOWN\Downarrow\rangle&=\left(|S\frac{1}{2},-\frac{1}{2}\rangle\right) = |s\rangle |\downarrow\rangle \label{eq:states_2}\\
|\UP\Uparrow\rangle&=\left(|P\frac{1}{2},\frac{1}{2}\rangle \right)= -\frac{|p_z\rangle |\uparrow\rangle+\left(|p_x\rangle + i|p_y\rangle\right) |\downarrow\rangle}{\sqrt{3}} \label{eq:states_3}  \\ 
|\UP\Downarrow\rangle&=\left(|P\frac{1}{2},-\frac{1}{2}\rangle\right) = \frac{|p_z\rangle |\downarrow\rangle-\left(|p_x\rangle - i|p_y\rangle\right) |\uparrow\rangle}{\sqrt{3}} \label{eq:states_4}
\end{align}  
where the left-hand-side introduces the `quasi-spin' notation convenient for our work. The right-hand-side follows the standard notation~\cite{Kane1966,Chuang1995} for the spin structure of a state [$|\downarrow\rangle$ and $|\uparrow\rangle$] and for the components of the Bloch functions [$|s\rangle$, $|p_x\rangle$, $|p_y\rangle$, $|p_z\rangle$]. For clarity, we also describe the states in terms of the total angular momentum and its projection in the parentheses. Note that in the companion paper~\cite{Volosniev2022}, for simplicity, we used a somewhat different notation, e.g., there $\Uparrow\uparrow$ was used instead of $\UP\Uparrow$. We do not employ this simplification in this paper.

In a periodic lattice, the atomic states form bands. In the basis given by Eqs.~(\ref{eq:states_1})-(\ref{eq:states_4}), the $s$- and $p$-type states form the valence and conduction band, respectively. To understand the corresponding physics, the standard approach is to construct an effective $\mathbf{k\cdot p}$ Hamiltonian that acts on these basis states. In the cubic phase, the resulting Hamiltonian in the matrix form acting on $\psi^T = \left(\UP\Uparrow,\, \UP\Downarrow,\, \DOWN\Uparrow,\, \DOWN\Downarrow\right)$ can be written down in the vicinity of the high-symmetry $R$-point of the Brillouin zone as follows ($\hbar=1$): 

\begin{equation}
H_k = 
\begin{pmatrix}
\frac{\Delta(k a)}{2} & 0                             & -2it k_z a                   & -2it k_- a \\
0                             & \frac{\Delta(ka)}{2} & -2it k_+ a                   & 2it k_z a  \\
2itk_z a                   & 2itk_- a                    & -\frac{\Delta(k a)}{2} & 0            \\
2it k_+ a                   & -2it k_z a                  & 0                              & -\frac{\Delta(k a)}{2}\\
\end{pmatrix}
\label{eq:matrixH}
\end{equation}

\noindent here $k_i$ is the momentum of the electron (without loss of generality, we assume that ${\bf k}\!=\!0$ corresponds to the $R$-point); $k_{\pm}=k_x\pm i k_y$; $a$ is the (cubic) lattice unit; $\Delta(ka)$ has the meaning of a $k$-dependent energy gap with a minimum at the $R$-point; $(i t)$ is the intra-orbital (s-p) overlap integral between the neighbouring sites~\cite{Becker2018}.

The $\mathbf{k\cdot p}$ method is a powerful tool with a general applicability~\cite{Kane1966, Chuang1995}. However it turns out that for the specific case of a cubic lead-halide perovskite it could be of advantage to take an alternative route, and construct an effective description of the low-energy physics from the allowed symmetries. To this end, we note that any operator $\hat O$~\footnote{Note that we shall use the `hat' over the letter to denote operators only when it is needed for clarity. We shall omit the `hat' when such an omission cannot cause any confusion.} that acts in the Hilbert space based upon the states in Eqs.~(\ref{eq:states_1})-(\ref{eq:states_4}), can be written as follows:
\begin{equation}
\hat O=\sum_{i,j,l} \mathcal{C}^l_i \mathcal{D}^l_j\tau^{\UP}_i\otimes \tau^{\Uparrow}_j,
\label{eq:O_general} 
\end{equation}
where $\tau_1,\tau_2,\tau_3$ is a set of the Pauli matrices and $\tau_0$ is the identity matrix. The subscript $\UP$ ($\Uparrow$) defines in a natural way the subset of the space where the matrices act. To simplify the notation, we shall define 
\begin{equation}
\tau_i\equiv \tau^{\UP}_i, \qquad \sigma_i\equiv \tau^{\Uparrow}_i.
\end{equation}
The expansion coefficients $\mathcal{C}_{i}^l$ and $\mathcal{D}_{j}^l$ can depend on the momentum of the particle as well as on the external electric, $\mathbf{E}$, and magnetic, $\mathbf{B}$, fields. If one assumes that the system is isotropic, then $\{\mathcal{C}_{1}^l,\mathcal{C}_{2}^l,\mathcal{C}_{3}^l\}$ and $\{\mathcal{D}_{1}^l,\mathcal{D}_{2}^l,\mathcal{D}_{3}^l\}$ should transform like vectors under the change of the system of coordinates. We shall assume cubic symmetry $O_h$ to model LHP. Therefore, in our work, the Hamiltonian becomes approximately isotropic only in the limit of low momenta.  In general, one expects an anisotropic optical response, see Sec.~\ref{sec:non_linear} for details.

Naturally, any effective Hamiltonian that describes the system can be written as Eq.~(\ref{eq:O_general}). Its most general form can be obtained by fixing time-reversal, and parity symmetries of the system. It is straightforward to show that the operators that implement these symmetries for the states in Eqs.~(\ref{eq:states_1})-(\ref{eq:states_4}) are
\begin{align}
\hat T= i\tau_3\otimes \sigma_2 \hat K,\\
\hat P=-\tau_3\otimes\sigma_0,
\end{align}
where $\hat K$ is the complex conjugation operator. 
The operators $\hat T$ and $\hat P$ commute with each other as they should.

Note that the time-reversal operator $\hat T$ acts in the standard way in the $\DOWN$ manifold
\begin{equation} 
\hat T|\DOWN\Uparrow\rangle=|\DOWN\Downarrow\rangle \qquad \hat T|\DOWN\Downarrow\rangle=-|\DOWN\Uparrow\rangle.
\end{equation}
However, there is an unconventional sign when $\hat T$ operates in the $\UP$ manifold 
\begin{equation}
\hat T|\UP\Uparrow\rangle=-|\UP\Downarrow\rangle \qquad \hat T|\UP\Downarrow\rangle=|\UP\Uparrow\rangle.
\end{equation}
Analogously, one can check that the parity operation also depends on the manifold.

\subsection{Effective Hamiltonian} 

To construct an effective Hamiltonian, let us first consider the simplest case of vanishing momentum ($\mathbf{k}=0$) and no external electromagnetic fields ($\mathbf{E}=0, \mathbf{B}=0$). In this case, there are only three commuting operators: 
$\hat T$, $\hat P$ and the identity operator, $\hat I$. The operator $\hat T$ is anti-unitary and cannot 
enter the Hamiltonian. Therefore, the Hamiltonian must be of the form
\begin{equation}
H(\mathbf{k}=\mathbf{E}=\mathbf{B}=0)=\alpha^{(1)}\tau_0\otimes\sigma_0+\alpha^{(2)}\tau_3\otimes\sigma_0,
\label{eq:Ham_zero}
\end{equation}
where the parameter $\alpha^{(1)}$ determines the offset of the energy~\footnote{In principle, this parameter can take into account the electric potential, and differences in the masses of electrons and holes. This possibility is not explored in this work.}. It will not be important in our study and can be set to zero. The physics of the parameter $\alpha^{(2)}$ is also clear; it fixes the gap between the $\UP$ and $\DOWN$ manifolds. One can conveniently write it as $\Delta/2$. For the APbBr$_3$ perovskites the value of the gap is known to be approximately 2~eV; see also below as well as Ref.~\cite{Kim2020} and references therein.

For a non-vanishing momentum and electromagnetic fields, one can add terms to Eq.~(\ref{eq:Ham_zero}). Indeed, in this case the coefficients $\mathcal{C}^l_i$ and  $\mathcal{D}^l_j$ can 
depend on $\mathbf{k}$, $\mathbf{E}$ and $\mathbf{B}$, and the most general of the Hamiltonian can be written as 
\begin{align}
H=&\alpha^{(1)}\tau_0\otimes\sigma_0+\alpha^{(2)}\tau_3\otimes\sigma_0 \nonumber + 
\sum_l \alpha^{(3)}_l\tau_1\otimes\sigma_l+& \\
& \sum_l \alpha^{(4)}_l \tau_2\otimes\sigma_l+\sum_l \alpha^{(5)}_l\tau_0\otimes\sigma_l+\sum_l \alpha^{(6)}_l\tau_3\otimes\sigma_l,&
\label{eq:Ham_non_zero_k_E}
\end{align}
where $\alpha^{(1)}$ and $\alpha^{(2)}$ are symmetric under the action of both $\hat T$ and $\hat P$; $\alpha^{(4)}_l$ is antisymmetric under $\hat T$ and $\hat P$; $\alpha^{(3)}_l$ is symmetric under $\hat T$, but antisymmetric under $\hat P$; $\alpha^{(5)}_l$ and $\alpha^{(6)}_l$ are antisymmetric under $\hat T$, but symmetric under $\hat P$. As before, we shall assume that $\alpha^{(1)}=0$.
Note that by definition $\hat T\mathbf{k}=\hat P\mathbf{k}=-\mathbf{k}$, $\hat T\mathbf{E}=-\hat P\mathbf{E}=\mathbf{E}$, and $\hat T\mathbf{B}=-\hat P\mathbf{B}=-\mathbf{B}$.

Assuming that the electromagnetic fields are weak, we can write (see also~\cite{Volosniev2022})
\begin{equation}
H=H_k+H_{E}+H_B,
\end{equation}
where $H_k$ determines the dispersion relation of the system. $H_{E}$ and $H_B$ determine coupling of LHP to electromagnetic fields beyond the minimal coupling to $H_k$.
Without loss of generality, we write the operator $H_k$ as
\begin{equation}
H_k=\frac{1}{2}\Delta(\tilde k)\tau_3\otimes \sigma_0 + 2 t  \tau_2 \otimes \sum_{l=1}^3\sigma_l S(\tilde k_l),
\end{equation} 
where $\tilde k_l=k_la-qa A_l $ ($q$ is the charge of a particle, $A_l$ is the vector potential in SI units) and $t$ is the hopping parameter. The even function $\Delta(x)$ and the odd function $S(x)$ define the momentum dependence of the electronic band. In what follows, we shall also write $\Delta(x)$ as 
\begin{equation}
\Delta(\tilde k)=\left(\Delta+t_3 \sum_{l=1}^3  C(\tilde k_l)\right),
\end{equation} 
where $C(x)$ is some even function. We assume that $C(x)\simeq x^2/2$ and $S(x)\simeq x$ in the limit $x\to 0$.  This natural assumption implies that in the limit $k\to 0$ the Hamiltonian $H_k$ corresponds to the model of Ref.~\cite{Becker2018}, see Eq.~(\ref{eq:matrixH}). 

For weak external fields, the operators $H_{E}$ and $H_B$ should have the following forms 
\begin{align}
\label{eq:H_E}
H_{E}=\mu\tau_1\otimes \sum_{l=1}^3 \sigma_l E_l,\\
H_{B}=(\mu_B^{(1)}\tau_0+\mu_B^{(2)}\tau_3)\otimes\sum_{l=1}^3 \sigma_l B_l,
\label{eq:H_B}
\end{align}
where $\mu$ controls 
the response of the medium to the external electric field (similar to the dipole moment). The parameters $\mu^{(1)}_B$ and $\mu^{(2)}_B$ determine the magnetic susceptibility. Note that the terms $[\vec k\times \vec E]$ and $[\vec k\times \vec B]$ have the symmetries of magnetic and electic fields, respectively. Indeed, $[\vec k\times \vec E]$ is a time-reversal-odd axial vector; $[\vec k\times \vec B]$ is a time-reversal-even polar vector. Therefore, they can be used in the corresponding places in Eqs.~(\ref{eq:H_E}) and~(\ref{eq:H_B}). However, these terms can be written as products of the basis coupling terms already presented in Eqs.~(\ref{eq:H_E}) and~(\ref{eq:H_B}); hence, they do not lead to any new functional dependence of observables on the parameters of the system, and are not important for our discussion. We leave their investigation to future studies.

Although, from the symmetry arguments
$\mu^{(1)}_B$ and $\mu^{(2)}_B$ can be arbitrary, the microscopic nature of the basis states imply that $\mu_B^{(1)}=-\mu_B^{(2)}$, which is explored in Ref.~\cite{Volosniev2022}. This condition corresponds to the fact that the Zeeman effect -- direct coupling of the magnetic field to $\uparrow$ and $\downarrow$ -- can occur only in the two lower states. Indeed, the Zeeman term does not act on the two higher states -- otherwise it would involve states outside the Hilbert space defined in Eqs.~(\ref{eq:states_1})-(\ref{eq:states_4}). For the sake of discussion, we shall treat $\mu^{(1)}_B$ and $\mu^{(2)}_B$ as independent quantities. This will help us to illustrate that the strength of the Faraday effect is given only by the first term in Eq.~(\ref{eq:H_B}), i.e., $\mu_B^{(1)}$, see Sec.~\ref{sec:faraday}.

As a summary of this section, the Hamiltonian in Eq.~(\ref{eq:Ham_non_zero_k_E}) is the low-energy description of the system assuming the Hilbert space from Eqs.~(\ref{eq:states_1})-(\ref{eq:states_4}), validating the phenomenological model presented in Ref.~\cite{Volosniev2022}. As expected, in the absence of external fields and for $k\to 0$, it coincides with the ${\bf k}\!\cdot\!{\bf p}$ Hamiltonian in Eq.~(\ref{eq:matrixH}). Strictly speaking, the effective Hamiltonian $H$ is limited to the vicinity of the band gap. However, most of the low-energy optical properties are determined by the transitions in this region, making the proposed effective model useful. Another merit of the presented phenomenological approach is that using similar symmetry arguments one can further amend the effective Hamiltonian with new degrees of freedom of a known symmetry.

\begin{figure}
\includegraphics[scale=0.4]{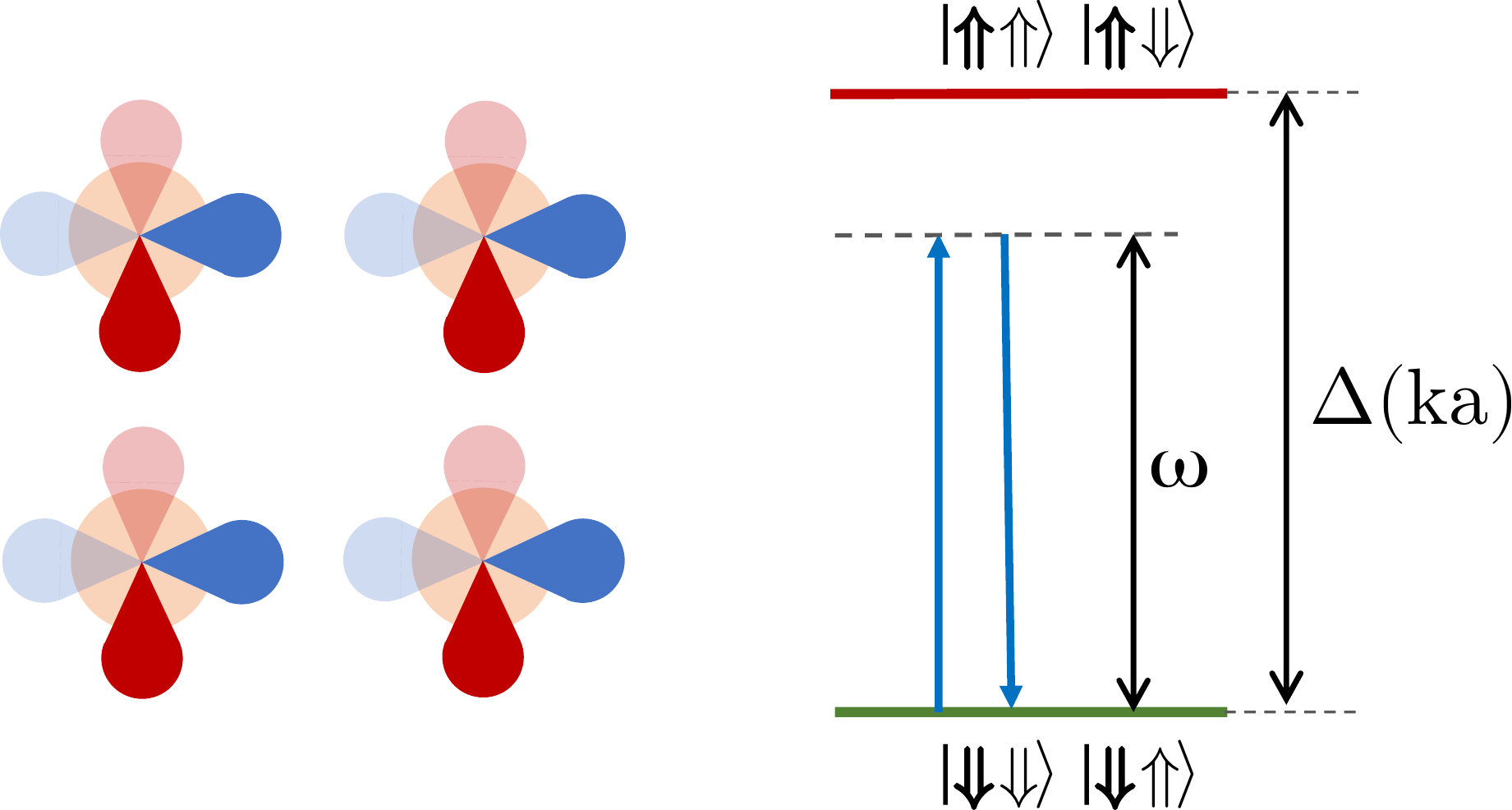}
\caption{ (left) Cartoon of the cubic lattice with $s$- and $p$-type orbitals. (right) The resonant band gap transition that contributes to the polarizability in Eq.~(\ref{eq:alpha_zz}). The vertical lines show the eigenstates of the Hamiltonian in Eq.~(\ref{eq:H0_for_alpha}) for $t=0$. They are separated by $\Delta(ka)$. The horizontal arrows show a transition to a virtual level driven by a photon with frequency~$\omega$.}.
\label{fig:levels}
\end{figure}

\section{Linear Optical Polarizability}
\label{sec:polarizability}

\subsection{General derivations}

\begin{figure}
\includegraphics[scale=0.25]{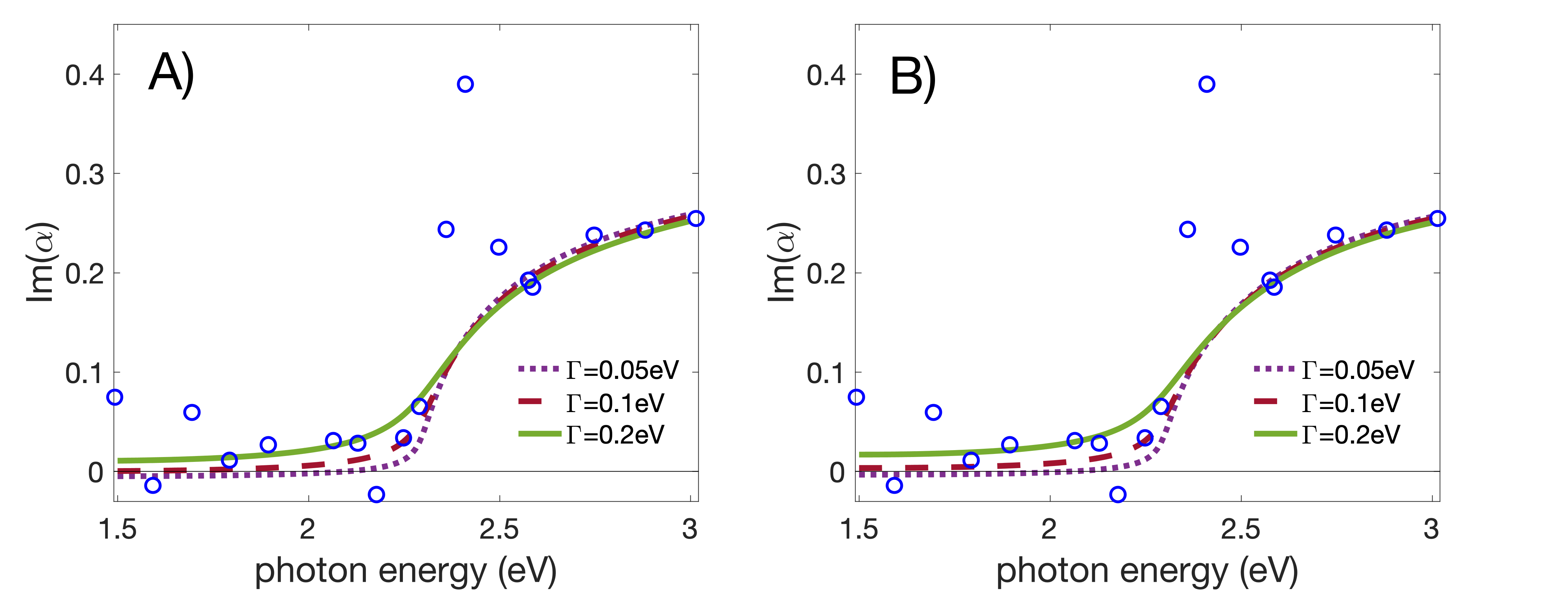}
\caption{Experimental (blue dots, Ref.~\cite{Volosniev2022}) and theoretical values for polarizability $\mathrm{Im}(\alpha_{zz})/\varepsilon_0$ of CH$_3$NH$_3$PbBr$_3$ as a function of the photon energy. Theoretical curves are calculated using Eq.~(\ref{eq:im_alpha_zz}) for different values of $\Gamma$.  Equation~(\ref{eq:C_S_1}) and $\mu=0.29 qa$ are used to produce panel A). Panel B) is for Eq.~(\ref{eq:C_S_2}) and $\mu=0.26 qa$. The other parameters that enter Eq.~(\ref{eq:im_alpha_zz}) are discussed in Sec.~\ref{subsec:experiment}.}
\label{fig:polarizability}
\end{figure}

As a first application of the effective Hamiltonian, we calculate the linear optical susceptibility in the vicinity of the energy gap. To this end, we follow the standard procedure~\cite{Boyd2008}.
First, we assume that the magnetic field is zero, and the electric field is weak ($E_l,A_l\to 0$). This allows us to write the Hamiltonian as
\begin{equation}
H\simeq H_0+H_P(t),
\label{eq:H_0plusH_P}
\end{equation}
where the time-independent part reads as
\begin{equation}
H_0=\frac{1}{2}\Delta(k a)\tau_3\otimes \sigma_0 + 2 t  \tau_2 \otimes \sum_{l=1}^3\sigma_l S(k_l a),
\label{eq:H0_for_alpha}
\end{equation}
and the time-dependent perturbation has the form
\begin{align}
\label{eq:HP_alpha}
H_P=&\mu\tau_1\otimes \sum_{l=1}^3\sigma_l E_l -& \\ &qa\sum_{l=1}^3 A_l\left(\frac{t_3}{2}C(k_l a)'\tau_3\otimes\sigma_0+2t\tau_2\otimes\sigma_l S(k_l a)'\right).&
\nonumber
\end{align}
Note that we have neglected the terms at the order of $A_l^2$ and beyond to be consistent with the derivation of the effective Hamiltonian (in particular of $H_E$ in Eq.~(\ref{eq:H_E})).  These terms are not immediately important for the discussion below, however they should be included to describe the optical properties at low frequencies (see Appendix~\ref{sec:appendix}).

To calculate the polarizability, we first use perturbation theory to calculate the change in the energy of the material in the approximation of slow fields. Then, we differentiate it with respect to the electric fields.  This procedure provides a general way for calculating susceptibilities of the systems that are described by the Hamiltonian $H$, see Appendix~\ref{sec:appendix}. 

Close to the band gap, the resonant transition dominates the optics (see Fig.~\ref{fig:levels}), and we derive in the leading order in $t/\Delta$ the polarizability of a unit volume 
\begin{equation}
\alpha_{zz}(\omega)=\int \frac{\mathrm{d}\mathbf{k}}{(2\pi\xi)^3}\frac{\left(\mu+2tS(k_z a)'\frac{qa}{\omega}\right)^2}{\Delta(k a)-\omega-i\Gamma/2},
\label{eq:alpha_zz}
\end{equation}
where $\omega$ is the frequency of light, $\Gamma$ is a phenomenological parameter to model non-unitary processes, such as decay of energy levels, and $\xi$ is a phenomenological geometric factor, see  Appendix~\ref{sec:appendix}. Note that Eq.~(\ref{eq:alpha_zz}) should be used only in the vicinity of the energy gap ($\omega\simeq \Delta$), in particular, it does not lead to a finite value of $\alpha_{zz}$ as $\omega\to 0$. The polarizability at low frequencies requires additional calculations as we discuss in Appendix~\ref{sec:appendix}.

The theoretical prediction of Eq.~(\ref{eq:alpha_zz}) can be connected to the refractive index via the Clausius–Mossotti relation (also known as the Lorentz–Lorenz law, see, e.g.,~\cite{FeynmanVol2,Boyd2008}) 
\begin{equation}
n^2=1+\frac{3 \alpha_{zz}}{3\varepsilon_0-\alpha_{zz}},
\end{equation}
where  $\varepsilon_{0}$ is the permittivity of free space. The integer $3$ is used here in approximation of an isotropic material; in general, it can also be used as a fitting parameter.
Assuming that the imaginary parts of $n$ and $\alpha_{zz}$ are small, we derive
\begin{align}
(\mathrm{Re} (n))^2\simeq 1+\frac{3\mathrm{Re}\alpha_{zz}}{3\varepsilon_0-\mathrm{Re}\alpha_{zz}},
\label{eq:RenFaraday}
\\
\mathrm{Im}(n)\simeq \frac{9\varepsilon_0\Im(\alpha_{zz})}{2\mathrm{Re}(n)[3\varepsilon_0-\mathrm{Re}(\alpha_{zz})]^2},
\end{align}
which leads to
\begin{equation}
 \frac{\Im(\alpha_{zz})}{\varepsilon_0}\simeq \frac{18\mathrm{Re}(n)\mathrm{Im}(n)}{\left[(\mathrm{Re}(n))^2+2\right]^2}.
\label{eq:im_alpha_zz}
\end{equation}
In our experimental set-up~\cite{Volosniev2022}, we can measure $\mathrm{Re}(n)$ and $\mathrm{Im}(n)$, which allows us to benchmark our theoretical calculations against experimental measurements. Note that  $\mathrm{Re}(n)\simeq 2$ for the considered parameters, which leads to 
$\Im(\alpha_{zz})\simeq \varepsilon_0 \mathrm{Im}(n)$. 

\subsection{Comparison to the experiment}
\label{subsec:experiment}

\begin{figure}
\includegraphics[scale=0.6]{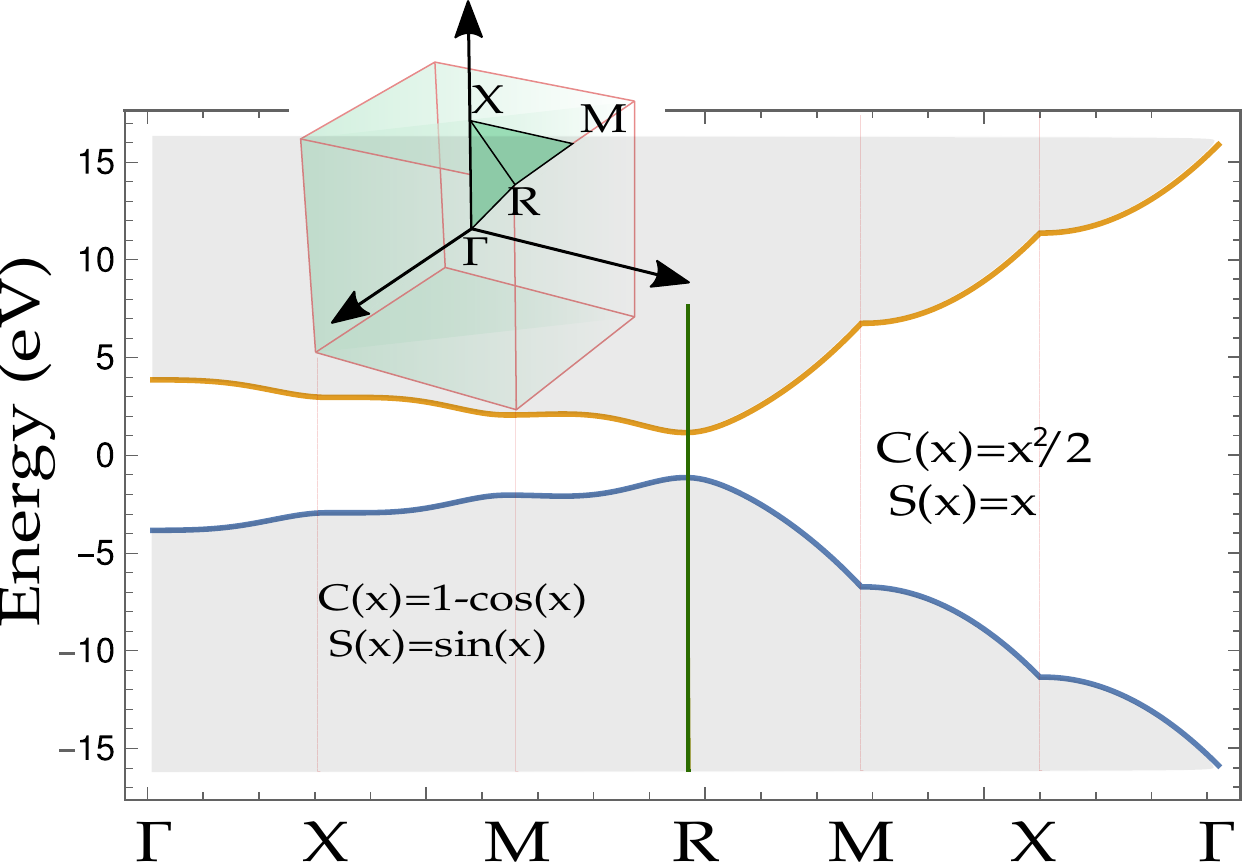}
\caption{The energy spectrum of the Hamiltonian without external fields, $\pm \sqrt{\frac{\Delta(ka)^2}{4}+4t^2\sum_l S(ka)^2}$, for Eqs.~(\ref{eq:C_S_1}) [to the left of the vertical (green) line] and~(\ref{eq:C_S_2}) [to the right of the vertical (green) line]. The positions in the Brillouin zone are depicted by $\Gamma$, $X$, $M$ and $R$ (see the inset). The depicted path $\Gamma\to X\to M\to R$ is motivated in part by the discussion of mirrors of bonding in Ref.~\cite{Goesten2018}.}
\label{fig:band_structure}
\end{figure}

Here, we use experimental values of $\mathrm{Re}(n)$ and $\mathrm{Im}(n)$ discussed in Ref.~\cite{Volosniev2022} to calculate the right-hand-side of Eq.~(\ref{eq:im_alpha_zz}). To estimate the left-hand-side, we work with our theoretical prediction presented in Eq.~(\ref{eq:alpha_zz}). 
To fix the parameters that enter Eq.~(\ref{eq:alpha_zz}), we rely on available experimental data~\cite{Pazhuk1981, Saidaminov2015, Jacobsson2016,Ng2018}, and the numerical data for CsPbBr$_3$~\cite{Becker2018}. 
These data motivate us to use $t=0.6$eV, $t_3=0.9$eV, $\Delta=2.3$eV, $a=0.586$nm (cf.~\cite{Volosniev2022}). To investigate the role of the functions $C(x)$ and $S(x)$, we employ two possible functional forms:
\begin{equation}
C(x)=1-\cos(x), \qquad S(x)=\sin(x),
\label{eq:C_S_1}
\end{equation} 
and
\begin{equation}
C(x)=\frac{x^2}{2}, \qquad S(x)=x.
\label{eq:C_S_2}
\end{equation}
As we show in Fig.~(\ref{fig:polarizability}), Eqs.~(\ref{eq:C_S_1}) and (\ref{eq:C_S_2}) lead to similar results for $\mathrm{Im}(\alpha_{zz})$, in spite of very different energy spectra (see Fig.~\ref{fig:band_structure}). This is a manifestation of the fact that only the behavior in the vicinity of $k\to 0$ is important for interpreting the present data.

The fit parameter $\mu$ is almost independent of $\Gamma$, assuming reasonable values of~$\Gamma$. Note that in the companion paper~\cite{Volosniev2022} we included also the fit to the exciton peak. 
We do not do it here, as an analysis of the exciton peak requires calculations beyond our single-body theoretical model; moreover, the value of $\mu$ is (almost) not sensitive to the inclusion of the exciton peak in the fitting procedure.  Another remark is that our model is capable of describing the shape and the amplitude of $\mathrm{Im}(\alpha_{zz})$ without electron-hole interactions.

\section{Faraday Effect}
\label{sec:faraday}

As the next application of our effective model, we consider the Faraday effect~\cite{Zommerfeld1954}, in which (linear) polarization of an electromagnetic wave 
is rotated in the presence of a magnetic field co-linear with propagation of light. 
First, we focus on a simple scenario without hopping terms ($t=t_3=0$). Then, we consider a general case.

\subsection{Faraday effect with $t=t_3=0$}
\begin{figure}
\includegraphics[scale=0.5]{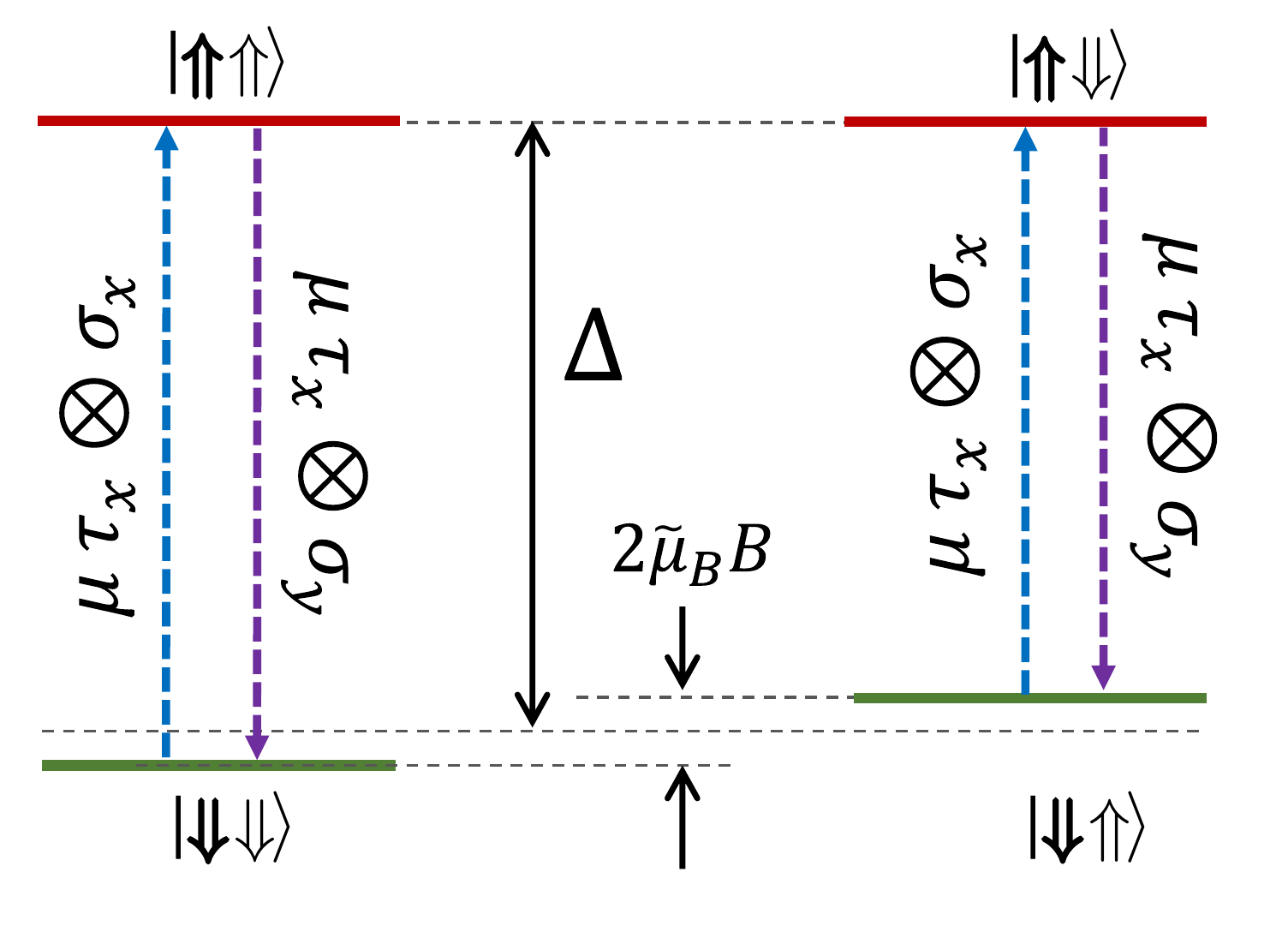}
\caption{Transitions that enter calculations of $\chi_{xy}$ for $t=t_3=0$, see Eq.~(\ref{eq:chi_xy_Faraday}). Here, to simplify the figure, we employ the assumption of the companion paper~\cite{Volosniev2022}: $\mu_B^{(1)}=-\mu_B^{(2)}=-\tilde \mu_B/2$. We do this without loss of generality, since only $\mu_B^{(1)}$ enters calculations.}.
\label{fig:levels_Faraday}
\end{figure}

To illustrate the origin of the Faraday effect in LHP, we calculate linear susceptibilities using the standard expression for the linear susceptibility (see Appendix~\ref{sec:appendix:Faraday})
\begin{align}
\chi_{xy}(\omega)\left[=-\chi_{yx}(\omega)\right]\simeq -\frac{i\mu^2 \omega}{\epsilon_0}\frac{8\mu^{(1)}_B \Delta }{(\Delta^2-\omega^2)^2}B,
\label{eq:chi_xy_Faraday}
\end{align}
and
\begin{align}
\chi_{xx}(\omega)\left[=\chi_{yy}(\omega)\right] \simeq \frac{2\mu^2}{\epsilon_0}\frac{\Delta}{\Delta^2-\omega^2}.
\label{eq:chi_xx_Faraday}
\end{align}
Note that only $\mu^{(1)}_B$ enters Eq.~(\ref{eq:chi_xy_Faraday}) implying that only the first term in Eq.~(\ref{eq:H_B}) contributes to $\chi_{xy}$. The term $\mu_B^{(2)}\tau_3\otimes\sum_{l=1}^3 \sigma_l B_l$ does not modify the energy differences between levels for the relevant transitions, see Fig.~\ref{fig:levels_Faraday}, and hence does not modify the susceptibility.

To show that Eqs.~(\ref{eq:chi_xy_Faraday}) and~(\ref{eq:chi_xx_Faraday}) lead to the Faraday effect, 
we consider the susceptibility matrix $\hat \chi$ written in the following form
\begin{equation}
\hat \chi(\omega)=\chi_{xx}(\omega)
\begin{pmatrix}
1& i \mathrm{Im}(\chi_{xy})/\chi_{xx}\\
-i\mathrm{Im}(\chi_{xy})/\chi_{xx} & 1
\end{pmatrix}.
\end{equation}
This form of $\hat \chi(\omega)$ conserves circular polarization of light 
\begin{equation}
\hat \chi\begin{pmatrix}1\\ \pm i\end{pmatrix}=\left(\chi_{xx}\mp \mathrm{Im}(\chi_{xy})\right)\begin{pmatrix}1\\ \pm i\end{pmatrix},
\end{equation}
allowing us to write the polarization vector as
\begin{equation}
P_{\pm}=\epsilon_0 (\chi_{xx}\mp\mathrm{Im}(\chi_{xy}))E_{\pm},
\end{equation}
where $\pm$ corresponds to the amplitudes of right- and left-polarized light.
Correspondingly, there are two indices of refraction
\begin{equation}
n_{\pm}=\sqrt{1+\chi_{xx}\mp\mathrm{Im}(\chi_{xy})},
\end{equation}
which lead to the Faraday effect.
The resulting Verdet constant enjoys the standard form for semiconductors~\cite{Boswarva1962}
\begin{equation}
V\equiv\frac{\omega}{2c}\frac{n_+-n_-}{B}\simeq-\frac{\omega}{2cB}\frac{\mathrm{Im}(\chi_{xy})}{\sqrt{1+\chi_{xx}}},
\end{equation}
where $\chi_{xy}$ should be taken from Eq.~(\ref{eq:chi_xy_Faraday}); instead of using the theoretical expression for $\chi_{xx}$ presented in Eq.~(\ref{eq:chi_xx_Faraday}), it is logical to use experimental data for the refractive index: $\sqrt{1+\chi_{xx}}\simeq \mathrm{Re}(n) $.

\subsection{General case}
In general, the Verdet coefficient assumes the form (see Appendix~\ref{sec:appendixB_2})
\begin{equation} 
V=\frac{\mu_B^{(1)}}{c\epsilon_0\mathrm{Re}(n)}\int\frac{\mathrm{d}\mathbf{k}}{(2\pi)^3}\frac{c_0+c_2\omega^2}{(\Delta(k a)^2-\omega^2)^2},
\label{eq:Verdet_general}
\end{equation}
where 
\begin{align}
c_0=&16\Delta(k a) t^2 S'(k_x a)S'(k_y a)(qa)^2& \nonumber\\
&+4t\mu\Delta(k a)^2(S'(k_x a)+S'(k_x a))qa,&
\end{align}
and 
\begin{equation}
c_2=4\Delta(k a)\mu^2+4tqa\mu(S' (k_x a)+S'(k_x a)).
\end{equation}
Note that $V$ is finite at $\omega=0$,
which is possible on general grounds if hopping is allowed~\cite{Bennett1965}.

The existence of the spin-electric term (i.e., $\mu\neq 0$) is crucial for explaining strong Faraday effect observed in the experiment~\cite{Sabatini2020, Volosniev2022}. Indeed, if $\mu=0$, then $c_2=0$ and the reduced Verdet coefficient $V(1-\omega^2/\Delta^2)^2$ has only weak frequency dependence, which contradicts experimental observations~\cite{Volosniev2022}. Moreover, even the overall increase of $V\cdot(1-\omega^2/\Delta^2)^2$ found experimentally cannot be reproduced.

We have observed that the Verdet coefficient at low frequencies is sensitive to the form of the functions 
$S(x)$ and $C(x)$. Therefore, one can provide only an order-of-magnitude estimate of $\mu_B^{(1)}$. 
To this end, we notice that if $t_3\to 0$, then  
\begin{equation} 
V=\frac{4\Delta\mu^2\omega^2\mu_B^{(1)}}{c\epsilon_0\mathrm{Re}(n)}\frac{1+\frac{t_3}{\Delta}F(\omega)}{(\Delta^2-\omega^2)^2},
\end{equation}
where $F(\omega)$ is some even function of $\omega$. Using this expression, we estimate $\mu_B^{(1)}=0.2\mu_B$, where $\mu_B$ is the Bohr magneton, see also Ref.~\cite{Volosniev2022}.

\section{Non-linear susceptibility}
\label{sec:non_linear}

The focus of the two previous sections was on linear response. 
The goal of this section is to demonstrate that the proposed model can also be 
used to calculate non-linear optical effects, which are important in LHP~\cite{Zhou2020}. 
Note that in centrosymmetric materials the even-order susceptibilities vanish. Therefore,
non-linearity in the leading order is given by the third-order susceptibility, $\chi^{(3)}$. 
We will show how to compute this quantity within our model with a particular focus on its geometrical properties.

As in Sec.~\ref{sec:polarizability}, we assume that the magnetic field is zero, the electric field is weak ($E_l,A_l\to 0$), and consider the Hamiltonian in the form
\begin{equation}
H\simeq \mathcal{H}_0+H_P(t),
\label{eq:H_0plusH_P_chi3}
\end{equation}
where the unperturbed Hamiltonian reads as follows
\begin{equation}
\mathcal{H}_0=\frac{1}{2}\Delta(k a)\tau_3\otimes \sigma_0.
\label{eq:H0_for_chi3}
\end{equation}
Note that for the sake of discussion we have neglected the term proportional to $t$ [cf. Eq.~(\ref{eq:H0_for_alpha})]. 
This omission implies that our results are correct only in the leading order in $t/\Delta$. 

We write the time-dependent perturbation from Eq.~(\ref{eq:HP_alpha}) in the form $H_P=H_P^{(1)}+H_P^{(2)}$,
where $H_P^{(1)}$ induces transitions between the eigenstates of $\mathcal{H}_0$: 
\begin{align}
\label{eq:HP1_chi3}
H_P^{(1)}=\mu\tau_1\otimes \sum_{l=1}^3\sigma_l E_l -2t qa\sum_{l=1}^3 A_l\tau_2\otimes\sigma_l S(k_l a)'.
\end{align}
and $H_P^{(2)}$ does not:
\begin{align}
\label{eq:HP2_chi3}
H_P^{(2)}=-\frac{qa t_3}{2}\sum_{l=1}^3 A_lC(k_l a)'\tau_3\otimes\sigma_0.
\end{align}
Since $H_P^{(2)}$ does not lead to any transitions, it can be neglected when dealing with linear response. 
However, this term is important for non-linear processes, and third-harmonic generation (THG) in particular. Note also that to be consistent with the derivation of the model, Eqs.~(\ref{eq:HP1_chi3}) and ~(\ref{eq:HP2_chi3}) are truncated at the linear order of $A_l$. The higher-order terms become important when describing the low-frequency limit, which is beyond the scope of this paper.

\subsection{THG with $H_P^{(2)}=0$}
Let us first consider the case in which $H_P=H_P^{(1)}$. 
Far from resonances the imaginary part of $\chi^{(3)}$ vanishes, and the third-order susceptibility reads as
\begin{align}
\label{eq:chi3_general}
\bar \chi_{ljih}^{(3)}&(\Omega,\omega,\omega',\omega'')= &\\&A \hat P_F\sum_{\nu,m,n}\frac{d_{g\nu}^l(-\Omega)d_{\nu n}^j(\omega)d_{nm}^i(\omega')d_{mg}^h(\omega'')}{(\varepsilon_{\nu g}+\Omega)(\varepsilon_{ng}+\omega'+\omega'')(\varepsilon_{mg}+\omega'')}&\nonumber,
\end{align}
where $\Omega=\omega+\omega'+\omega''$; $A$ is a constant that depends on the density of charges and the system of units,
$\hat P_F$ is an operator that produces 24 terms by permuting $-\Omega,\omega,\omega'$ and $\omega''$ together with the corresponding Cartesian indices. The bar symbol ($\bar \chi$) is used to specify that $H_P=H_P^{(1)}$;  the index $g$ refers to the ground state; the sum is over all possible intermediate states; a summation over all momenta is implicitly implied. The energy difference is $\varepsilon_{ng}=\varepsilon_n-\varepsilon_g$,
where $\varepsilon_n=\pm\Delta(k a)/2$ is the eigenenergy of the unperturbed Hamiltonian. 
The matrix elements are defined as 
\begin{equation}
d_{\nu n}^j(\omega)=\langle\nu|\mu\tau_1\otimes\sigma_j-2 i \frac{qa}{\omega} t  \tau_2 \otimes \sigma_j S(k_j a)'|n\rangle.
\end{equation}
Note that in comparison to the standard expression~\cite{Boyd2008}, the matrix elements depend on frequencies, which is taken into account in our calculations, see also App.~\ref{app:chi3}. 

The expression in Eq.~(\ref{eq:chi3_general}) can be simplified
\begin{align}
\label{eq:chi3_general_simplified}
\bar \chi_{ljih}^{(3)}&(\Omega,\omega,\omega',\omega'')= &\\&A\hat P_F\int\frac{\mathrm{d}\mathbf{k}}{(2\pi)^3} \frac{f_l(\Omega)f_j(\omega)f_i(-\omega')f_h(\omega'')\langle\sigma_l\sigma_j\sigma_i\sigma_h\rangle_{\Uparrow}}{(\Delta(ka)+\Omega)(\omega'+\omega'')(\Delta(ka)+\omega'')}\nonumber,&
\end{align}
where $f_i(\omega)=\mu-2qatS(k_ia)'/\omega$, and $\langle\sigma_l\sigma_j\sigma_i\sigma_h\rangle_{\Uparrow}$ is the average over the degenerate ground state. It is easy to show that if $t=0$, then 
\begin{equation}
\label{eq:chi3_isotropic}
\bar \chi_{xxxx}^{(3)}=\bar \chi_{xxyy}^{(3)}+\bar \chi_{xyyx}^{(3)}+\bar \chi_{xyxy}^{(3)},
\end{equation}
which is a manifestation of the isotropic nature of the band structure at $k=0$. However, 
there can be a substantial anisotropic effect when the parameter $t$ is non-vanishing. 
To demonstrate this, we consider third-harmonic generation (THG), i.e., we consider 
$\bar \chi^{(3)}(3\omega,\omega,\omega,\omega)$.  

\begin{figure}
\includegraphics[scale=0.3]{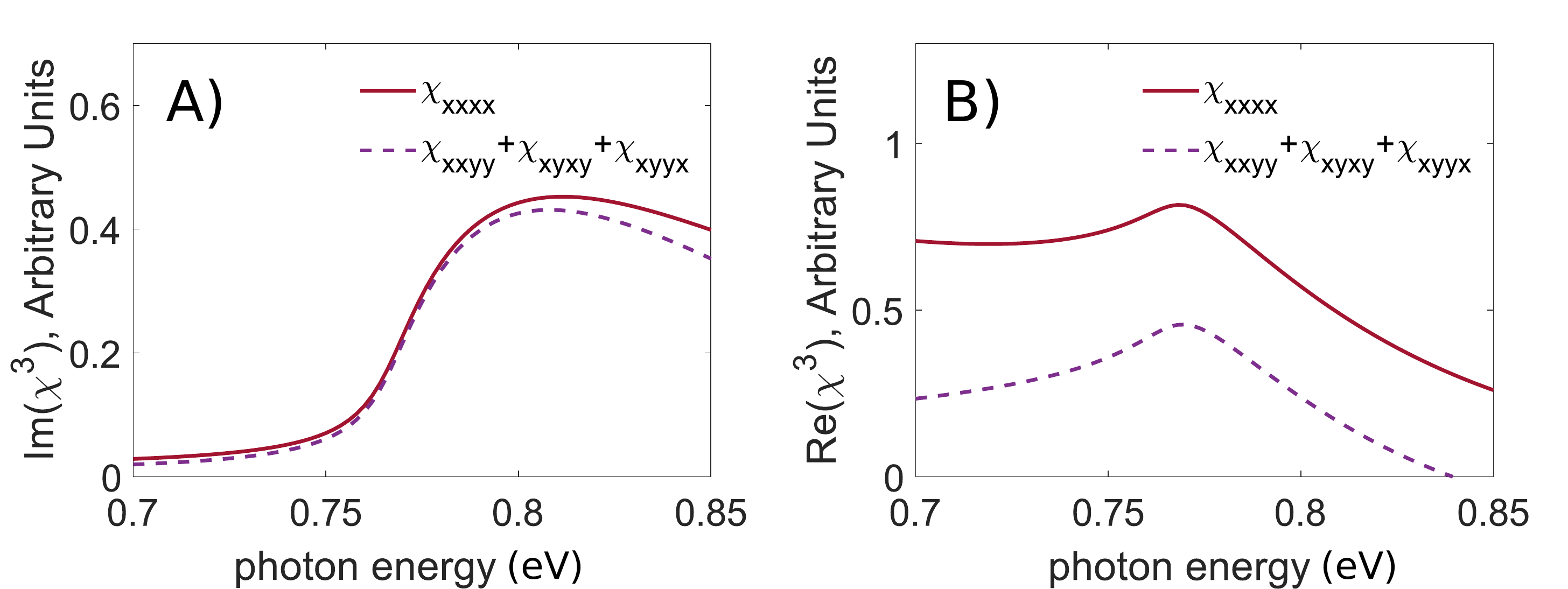}
\caption{The third-order susceptibility in the vicinity of a three-photon resonance transition. The left (right) panel shows the imaginary (real) part of $\chi^3$. 
In each panel, the upper curve presents $\chi_{xxxx}^{(3)}$, and the lower curve is for $\chi_{xyxy}^{(3)}+\chi_{xyyx}^{(3)}+\chi_{xxyy}^{(3)}$. The difference between the upper and lower curves signals that the optical response of LHP is anisotropic. To produce the data in the figure, the parameters presented in Sec.~\ref{subsec:experiment} were used; $\Gamma=0.05$eV. The band structure is defined by Eq.~(\ref{eq:C_S_1}). }.
\label{fig:absorption_chi3}
\end{figure}

The dominant contribution to this quantity is given by the terms in Eq.~(\ref{eq:chi3_general_simplified}) which contain in the denominator $(\Delta(ka)-3\omega)$. In general, there are six terms of this type (see App.~\ref{app:chi3}). However, they are highly symmetric, which allows us to write compact expressions:
\begin{align}
\bar \chi_{xxxx}^{(3)}\simeq  \frac{A}{\omega}\int\frac{\mathrm{d}\mathbf{k}}{(2\pi)^3} \frac{-3f_x(-3\omega)f_x(\omega)f_x(-\omega)^2}{(\Delta(ka)-\omega)(\Delta(ka)-3\omega - i\frac{\Gamma}{2})},
\nonumber
\end{align}
where we have introduced $\Gamma$ to regularize the integral for $\omega>\Delta/3$. 
The off-diagonal susceptibilities are equal to each other, i.e., $\bar \chi_{xyxy}^{(3)}=\bar \chi_{xyyx}^{(3)}=\bar \chi_{xxyy}^{(3)}$:
\begin{align}
\bar \chi_{xxyy}^{(3)}=\frac{A}{\omega}\int\frac{\mathrm{d}\mathbf{k}}{(2\pi)^3} \frac{f_x(-3\omega)f_x(\omega)f_y(-\omega)f_y(-\omega)}{(\Delta(ka)-\omega)(\Delta(ka)-3\omega-i\frac{\Gamma}{2})}\nonumber\\-\frac{2 A}{\omega}\int\frac{\mathrm{d}\mathbf{k}}{(2\pi)^3} \frac{f_x(-3\omega)f_x(-\omega)f_y(\omega)f_y(-\omega)}{(\Delta(ka)-\omega)(\Delta(ka)-3\omega-i\frac{\Gamma}{2})}.\nonumber
\end{align}
Note that if $t=0$, then the response is isotropic in agreement with Eq.~(\ref{eq:chi3_isotropic}).

\subsection{General Case}
As mentioned above, $H_P^{(2)}$ becomes important for non-linear processes. In particular, the term $H_P^{(2)}$ leads to additional terms in the third-order susceptibility:
$\chi_3=\bar \chi_3+{\bar{\bar\chi}}_3$, where ${\bar{\bar\chi}}_3$ can be calculated following the procedure described above. For third-harmonic generation, we derive   
\begin{align}
{\bar{\bar\chi}}_{xxxx}^{(3)}=A\int\frac{\mathrm{d}\mathbf{k}}{(2\pi)^3}\left[\frac{qa t_3 C(k_x a)'}{2\omega}\right]^2 \frac{\mathcal{F} f_x(-\omega)f_x(-3\omega)}{\Delta(k a)-3\omega-i\frac{\Gamma}{2}},
\nonumber
\end{align}
and
\begin{align}
{\bar{\bar\chi}}_{xxyy}^{(3)}=\frac{A}{3}\int\frac{\mathrm{d}\mathbf{k}}{(2\pi)^3}\left[\frac{qa t_3 C(k_y a)'}{2\omega}\right]^2 \frac{\mathcal{F} f_x(-\omega)f_x(-3\omega)}{\Delta(k a)-3\omega-i\frac{\Gamma}{2}},
\nonumber
\end{align}
where 
\begin{equation}
\mathcal{F}=-3\frac{\Delta(k a)(\Delta(k a)-\omega)+2\omega^2}{\omega^2(\Delta(ka)-\omega)(\Delta(ka)-2\omega)}.
\nonumber
\end{equation}
Again, it is easy to see that the responce is only anisotropic if $t\neq 0$. 
To illustrate the anisotropy, we first calculate the imaginary part of $\chi^{(3)}$, see Fig.~\ref{fig:absorption_chi3}~a). We see that for the photon energies that are above the threshold the response is anisotropic. This is expected from the discussion above: When the light starts to probe the band structure with $k>0$, the responce includes the hopping term, which leads to a directional dependence of $\chi^3$. The real part of the third-order susceptibility always features anisotropy, see Fig.~\ref{fig:absorption_chi3}~b). Even in the vicinity of the three-photon resonance, i.e., when the response is the most isotropic, we have $\chi_{xxxx}\simeq 2(\chi_{xxyy}+\chi_{xyxy}+\chi_{xyyx})$. 
Therefore, one expects to see a strong directional dependence of third-harmonic generation.

Finally, we note that the anisotropy strongly depends on the band structure. For example, it is easy to show that Eq.~(\ref{eq:C_S_2}) leads to isotropic response, which can be used to experimentally distinguish between the band structures presented in Fig.~\ref{fig:band_structure}. As could be expected, the non-linear susceptibility contains information about the energy spectrum across different momenta, which can be used to discover properties of the material that are beyond reach of linear response.

\section{Summary and Outlook}
\label{sec:summary}

In this paper, which accompanies~\cite{Volosniev2022}, we introduced a natural extension of the $\mathbf{k\cdot p}$ Hamiltonian (see Eq.~(\ref{eq:matrixH})) that allows one to study optoelectronic phenomena in LHP in the near-infrared range
beyond minimal coupling substitution. In our study, we first used general symmetry constraints such as time-reversal and parity symmetry to identify possible terms in the effective model of LHP  in weak electromagnetic fields, see~Eq.~(\ref{eq:Ham_non_zero_k_E}). 

To test the model, we computed the linear optical polarizability and the Verdet constant. Comparison of our theoretical results to experimental data allowed us to confirm the necessity of the spin-electric and Zeeman terms, which go beyond those in the minimally coupled $\mathbf{k\cdot p}$ Hamiltonian. Furthermore, we used the experimental data to fix the introduced phenomenological parameters. 

 To illustrate a general applicability of our model to optical phenomena in LHP, we calculated
the third-order linear susceptibility. We argued that the spin-electric term can induce only isotropic optical response due to the symmetry of the band structure. At the same time, the hopping term can yield significant directional dependence. 
To demonstrate this, we calculated anisotropy in third-harmonic generation, see Fig.~\ref{fig:absorption_chi3}.  
These theoretical calculations can be confirmed experimentally (not discussed here), which will be the subject of a follow-up publication. Our findings pave the way for using anisotropy of optical response to learn about the band structure and microscopic physics of LHP.

The proposed effective Hamiltonian suggests a number of phenomena, such as Rashba-type splitting of energy levels and axion-type physics. These effects, which are briefly discussed in Ref.~\cite{Volosniev2022}, require further theoretical and experimental investigations. Furthermore, the presented phenomenological approach suggests a framework for symmetry-based inclusion of terms into effective models of LHP. In particular, it paves the way for studying the coupling to other relevant degrees of freedom (for example, given by the lattice). Investigation of the effect of the spin-electric term on excitons is another research direction that naturally follows from our work.

\begin{acknowledgments}
We thank Maksym Serbyn, Areg Ghazaryan and Nuh Gedik for useful discussions; M.L. acknowledges support  by the European Research Council (ERC) Starting
Grant No. 801770 (ANGULON).
\end{acknowledgments}

\appendix

\begin{widetext}
\section{Polarizability of the medium}
\label{sec:appendix}

Here, we work with the Hamiltonian from Eq.~(\ref{eq:H_0plusH_P}), $H=H_0+H_P(t)$, and derive the polarizability of the medium presented in Eq.~(\ref{eq:alpha_zz}). Let us now write time-independent eigenstates of $H_0$ as $|m\rangle$. 
A time-evolved eigenstate can be derived using first order perturbation theory
\begin{equation}
|\psi_n(t)\rangle = |n\rangle e^{-i \varepsilon_n t}-\sum_{m,\omega}\langle m|H_P(\omega)|n\rangle \frac{e^{-i\varepsilon_n t+i\omega t}}{\varepsilon_m-\varepsilon_n+\omega-i\Gamma/2}|m\rangle,
\end{equation}
where we have introduced decay via $\Gamma$, and used 
\begin{equation}
H_P(t)\equiv \sum_\omega H_P(\omega)e^{i\omega t}=\sum_{\omega,l} \tilde H^{l}_P(\omega) E^l(\omega)e^{i\omega t}.
\end{equation} 
Assuming that the perturbation is given by Eq.~(\ref{eq:HP_alpha}), the quantity $\tilde H^{l}_P(\omega)$ takes the form:
\begin{equation}
\tilde H^{l}_P(\omega)=\mu\tau_1\otimes\sigma_l-2 i \frac{qa}{\omega} t  \tau_2 \otimes \sigma_l S(k_l a)'   -
i \frac{qa}{\omega} \frac{t_3}{2}\tau_3\otimes \sigma_0 C(k_l a)',
\label{eq:HlPO}
\end{equation}
where we have used that $\mathbf{E}=-\frac{\partial \mathbf{A}}{\partial t}$.

To calculate linear susceptibility, we calculate the terms in the expectation value of the Hamiltonian that contain $E_l^2$ terms and have a finite value of time average:
\begin{equation}
\frac{\langle \psi_n(t)| H |\psi_n(t)\rangle}{\langle \psi_n(t)| \psi_n(t)\rangle} \to -\sum_{m,\omega,\omega'}\frac{\langle n|H_P(\omega')|m\rangle\langle m|H_P(\omega)|n\rangle}{\varepsilon_m-\varepsilon_n+\omega-i\Gamma/2}e^{i\omega t + i\omega't}-\sum_{m,\omega,\omega'}\frac{\langle m|H_P(\omega')|n\rangle(\langle m|H_P(\omega)|n\rangle)^*}{\varepsilon_m-\varepsilon_n+\omega+i\Gamma/2}e^{-i\omega t + i\omega't}.
\end{equation}
We average these terms over time and derive an expression of {\it the energy of the material} in the external field suitable for our calculations
\begin{equation}
\frac{\langle \psi_n(t)| H |\psi_n(t)\rangle}{\langle \psi_n(t)| \psi_n(t)\rangle}\bigg|_T\to \mathcal{E}=-\sum_{m,\omega}\frac{\langle n|H_P(\omega)|m\rangle\langle m|H_P(-\omega)|n\rangle}{\varepsilon_m-\varepsilon_n-\omega-i\Gamma/2}-\sum_{m,\omega}\frac{\langle m|H_P(\omega)|n\rangle(\langle m|H_P(\omega)|n\rangle)^*}{\varepsilon_m-\varepsilon_n+\omega+i\Gamma/2}.
\end{equation}
If we differentiate this energy with respect to $E_i(\omega)^*$ and $E_{j}(\omega)$, we derive the polarizability of the medium for a given value of the momentum $k$ 
\begin{align}
\alpha^{(k)}_{ij}(\omega)=\sum_{m}\bigg(\frac{\langle n|\tilde H^j_P(\omega)|m\rangle\langle m|\tilde H^i_P(-\omega)|n\rangle}{\varepsilon_m-\varepsilon_n-\omega-i\Gamma/2}+\frac{(\langle m|\tilde H^i_P(\omega)|n\rangle)^*\langle m|\tilde H^j_P(\omega)|n\rangle}{\varepsilon_m-\varepsilon_n+\omega+i\Gamma/2}\bigg),
\label{eq:alpha_k_ij}
\end{align}
which is a logical extension of textbook results (see, e.g.,~\cite{Boyd2008}) to our problem. The index $n$ refers to the ground state, which is double degenerate. Therefore, we need to average over the `spin' degree of freedom, $\Downarrow$. 

 It is straightforward to calculate the expression in Eq.~(\ref{eq:alpha_k_ij}) exactly. However, the resulting expression is cumbersome and does not provide physical insight. Therefore, we use $t=0$ in $H_0$, which is a natural assumption, since $t$ is much smaller than $\Delta$. Note that the last term in Eq.~(\ref{eq:HlPO}) does not induce any transitions within this approximation scheme. Hence, it can be neglected for calculations of $\alpha$ in the vicinity of the band gap transition, and we derive 
\begin{align}
\alpha^{(k)}_{ii}(\omega)=\frac{(\mu+2\frac{qa}{\omega} t S(k_i a)')^2}{\Delta(k a)-\omega-i\Gamma/2}+\frac{(\mu-2\frac{qa}{\omega} t S(k_i a)')^2}{\Delta(k a)+\omega+i\Gamma/2}.
\label{eq:alpha_k_ii}
\end{align}
Here, the first part is resonant in the vicinity of the band gap transition. It was used to derive 
Eq.~(\ref{eq:alpha_zz}). The phenomenological geometric parameter $\xi$ that enters Eq.~(\ref{eq:alpha_zz}) determines the size of the polarizable unit. Its meaning can be most easily understood by considering $t=t_3=0$. In this case, $\alpha_{zz}=\tilde n\frac{\mu^2}{\Delta-\omega-i\Gamma/2}$, where $\tilde n$ is the density of polarizable units. In our calculations, we use $\xi\simeq 2.2$, which reproduces the data well. Note that the value of $\xi$ can affect the value of $\mu$ obtained in the fitting. However, we have checked that the precise value of $\xi$ is not important for our main conclusions.

Finally, note that the expression in Eq.~(\ref{eq:alpha_k_ii}) is valid only in the vicinity of the band gap transition, and should be modified otherwise. For example, to have a meaningful expression in the limit $\omega\to 0$, one should include higher orders of $A_l$ in the expansion of $S(\tilde k)$, which leads to
\begin{align}
\alpha^{(k)}_{ii}(\omega\to 0)=\frac{(\mu+2\frac{qa}{\omega} t S(k_i a)')^2}{\Delta(ka)-\omega}+\frac{(\mu-2\frac{qa}{\omega} t S(k_i a)')^2}{\Delta(k a)+\omega} + \frac{8t^2S(k_i a)S''(k_i a)}{\Delta}\left(\frac{q a}{\omega}\right)^2.
\label{eq:alpha_k_ii_omega_0}
\end{align}
Therefore, for $\omega=0$, we have
\begin{align}
\alpha_{ii}(\omega=0)=\int\frac{\mathrm{d}\mathbf{k}}{(2\pi)^3}\left[\frac{2\mu^2}{\Delta(ka)}+\frac{8qat\mu S(k_i a)'}{\Delta(k a)^2}\right].
\end{align}

\section{Faraday Effect}
\label{sec:appendix:Faraday}

\subsection{Derivation of Eqs.~(\ref{eq:chi_xy_Faraday}) and~(\ref{eq:chi_xx_Faraday})}
For $t=t_3=0$, the Hamiltonian of the system reads as 
\begin{equation}
H=\frac{\Delta}{2}\tau_3\otimes \sigma_0  + (\mu_B^{(1)}\tau_0+\mu_B^{(2)}\tau_3)\otimes\sum_{l=1}^3 \sigma_l B_l + \mu\tau_1\otimes \sum_{l=1}^3 \sigma_l E_l.
\end{equation} 
In the derivation of the susceptibility, the last term should be considered as perturbation of the form $\left(-\sum \hat d_l E_l\right)$. The expression of $\chi_{ij}$ far from resonances in this case is~\cite{Boyd2008} 
\begin{equation}
\chi_{ij}(\omega)=\frac{\mathcal{N}}{\epsilon_0}\sum_m\left(\frac{d_{nm}^i d_{mn}^j}{\varepsilon_m-\varepsilon_n+\omega}+\frac{d_{nm}^jd_{mn}^i}{\varepsilon_m-\varepsilon_n-\omega}\right),
\label{eq:chi_ij_Faraday}
\end{equation}
where $d_{nm}^i=-\mu \langle n|\tau_1\otimes \sigma_i|m\rangle$ (as before, $n$ refers to the ground state), and $\mathcal{N}$ implies either integration over the momentum or the density of atoms. Without loss of generality we shall omit this symbol. The sum in Eq.~(\ref{eq:chi_ij_Faraday}) is over all possible states, $m$.

Assuming that the $B$-field is along the $z$-axis that is determined by the direction of light propagation,
we derive
\begin{align}
\chi_{ij}(\omega)=\frac{\mu^2}{2\epsilon_0}\left[\frac{\langle \DOWN\Downarrow| \sigma_i \sigma_j| \DOWN\Downarrow\rangle}{\Delta+2\mu^{(1)}_B B+\omega}+\frac{\langle \DOWN\Downarrow| \sigma_j \sigma_i| \DOWN\Downarrow\rangle}{\Delta+2\mu^{(1)}_B B-\omega}\right] +\frac{\mu^2}{2\epsilon_0}\left[\frac{\langle \DOWN\Uparrow| \sigma_i \sigma_j| \DOWN\Uparrow\rangle}{\Delta-2\mu^{(1)}_B B+\omega}+\frac{\langle \DOWN\Uparrow| \sigma_j \sigma_i| \DOWN\Uparrow\rangle}{\Delta-2\mu^{(1)}_B B-\omega}\right].
\end{align}
Equations~(\ref{eq:chi_xy_Faraday}) and~(\ref{eq:chi_xx_Faraday}) now follow, for example,
\begin{align}
\chi_{xy}(\omega)=\frac{i \omega \mu^2}{\epsilon_0}\left[\frac{1}{(\Delta+2\mu^{(1)}_B B)^2-\omega^2}-\frac{1}{(\Delta-2\mu^{(1)}_B B)^2-\omega^2}\right] \simeq -\frac{i\omega\mu^2}{\epsilon_0}\frac{8\mu^{(1)}_B B \Delta}{(\Delta^2-\omega^2)^2}.
\end{align}

\subsection{Derivation of Eq.~(\ref{eq:Verdet_general})}
\label{sec:appendixB_2}
To derive the Verdet coefficient, we calculate the linear susceptibility. To this end, we follow the routine discussed in App.~\ref{sec:appendix}, 
i.e., we consider the Hamiltonian as the sum, $H=H_0+H_P$, where
\begin{equation}
H_0=\frac{1}{2}\Delta(k a)\tau_3\otimes \sigma_0 + 2 t  \tau_2 \otimes \sum_{l=1}^3\sigma_l S(k_l a) + (\mu_B^{(1)}\tau_0+\mu_B^{(2)}\tau_3)\otimes\sum_{l=1}^3 \sigma_l B_l,
\label{eq:H0_AppendixB}
\end{equation}
and the time-dependent perturbation has the form
\begin{align}
H_P=\mu\tau_1\otimes \sum_{l=1}^3\sigma_l E_l -qa\sum_{l=1}^3 A_l\left(\frac{t_3}{2}C(k_l a)'\tau_3\otimes\sigma_0+2t\tau_2\otimes\sigma_l S(k_l a)'\right).
\label{eq:HP_AppendixB}
\end{align}

According to App.~\ref{sec:appendix}, the linear susceptibility far from resonances (i.e., $\Gamma=0$) reads as 
\begin{align}
\chi_{ij}(\omega)=\frac{1}{\epsilon_0}\int \frac{\mathrm{d}\mathbf{k}}{(2\pi)^3} \sum_{m}\bigg(\frac{\langle n|\tilde H^j_P(\omega)|m\rangle\langle m|\tilde H^i_P(-\omega)|n\rangle}{\varepsilon_m-\varepsilon_n-\omega}+\frac{(\langle m|\tilde H^i_P(\omega)|n\rangle)^*\langle m|\tilde H^j_P(\omega)|n\rangle}{\varepsilon_m-\varepsilon_n+\omega}\bigg),
\end{align}
where $\tilde H^i_P$ is defined in Eq.~(\ref{eq:HlPO}). The states $m$ and $n$ are eigenstates of $H_0$
in Eq.~(\ref{eq:H0_AppendixB}). This form of $\chi_{ij}$ will lead to two contributions to the Verdet coefficients: the first contribution is due to the change in the energy levels due to $B$ (also sometimes called the diamagnetic part), and the second one is due to the change of the eigenvectors~\cite{Bennett1965}. We focus on the dominant first part, which can be written as 
\begin{align}
\chi_{xy}(\omega)=\frac{1}{2\epsilon_0}\int\frac{\mathrm{d}\mathbf{k}}{(2\pi)^3} \sum_{m}\left(\frac{\langle \DOWN\Downarrow|\tilde H^y_P(\omega)|m\rangle\langle m|\tilde H^x_P(-\omega)|\DOWN\Downarrow\rangle}{\varepsilon_m-\varepsilon_{\DOWN\Downarrow}-\omega}+\frac{\langle \DOWN\Downarrow|\tilde H^x_P(-\omega)|m\rangle\langle m|\tilde H^y_P(\omega)|\DOWN\Downarrow\rangle}{\varepsilon_m-\varepsilon_{\DOWN\Downarrow}+\omega}\right)+\nonumber \\
\frac{1}{2\epsilon_0}\int\frac{\mathrm{d}\mathbf{k}}{(2\pi)^3}\sum_{m}\left(\frac{\langle \DOWN\Uparrow|\tilde H^y_P(\omega)|m\rangle\langle m|\tilde H^x_P(-\omega)|\DOWN\Uparrow\rangle}{\varepsilon_m-\varepsilon_{\DOWN\Uparrow}-\omega}+\frac{\langle \DOWN\Uparrow|\tilde H^x_P(-\omega)|m\rangle\langle m|\tilde H^y_P(\omega)|\DOWN\Uparrow\rangle}{\varepsilon_m-\varepsilon_{\DOWN\Uparrow}+\omega}\right),
\end{align}
where we average over the ground-state manifold.
Note that the $t_3$-term in Eq.~(\ref{eq:HP_AppendixB}) does not induce any transitions between states, and therefore does not contribute to the Verdet coefficient. Therefore, we can set $t_3=0$. The resulting expression for $\chi_{xy}$ reads 
\begin{align}
\chi_{xy}\left[=-\chi_{yx}\right]=\frac{i}{2\epsilon_0}\int\frac{\mathrm{d}\mathbf{k}}{(2\pi)^3}\left(\frac{4 t^2 S'(k_x a)S'(k_y a)\left(\frac{qa}{\omega}\right)^2+\mu^2+2\mu t(S'(k_x a)+S'(k_y a))\frac{qa}{\omega}}{\Delta(k a)+2\mu_B^{(1)}B-\omega}\right)+\nonumber \\ 
\frac{i}{2\epsilon_0}\int\frac{\mathrm{d}\mathbf{k}}{(2\pi)^3}\left(\frac{-4 t^2S'(k_x a)S'(k_y a)\left(\frac{qa}{\omega}\right)^2-\mu^2+2\mu t(S'(k_x a)+S'(k_y a))\frac{qa}{\omega}}{\Delta(k a)+2\mu_B^{(1)}B+\omega}\right)+ \nonumber \\ \frac{i}{2\epsilon_0}\int\frac{\mathrm{d}\mathbf{k}}{(2\pi)^3}\left(\frac{-4 t^2 S'(k_x a)S'(k_y a)\left(\frac{qa}{\omega}\right)^2-\mu^2-2\mu t(S'(k_x a)+S'(k_y a))\frac{qa}{\omega}}{\Delta(k a)-2\mu_B^{(1)}B-\omega}\right)+\nonumber \\\frac{i}{2\epsilon_0}\int\frac{\mathrm{d}\mathbf{k}}{(2\pi)^3}\left(\frac{4 t^2S'(k_x a)S'(k_y a)\left(\frac{qa}{\omega}\right)^2+\mu^2-2t\mu(S'(k_x a)+S'(k_y a))\frac{qa}{\omega}}{\Delta(k a)-2\mu_B^{(1)}B+\omega}\right).
\end{align}
This expression can be re-written as
\begin{equation} 
\chi_{xy}=\frac{-2i\mu_B^{(1)} B}{\epsilon_0}\int\frac{\mathrm{d}\mathbf{k}}{(2\pi)^3}\frac{4\omega\Delta(k a)\left(4 t^2S'(k_x a)S'(k_y a)\left(\frac{qa}{\omega}\right)^2+\mu^2\right)+4t\mu(\Delta(k a)^2+\omega^2)(S'(k_x a)+S'(k_x a))\frac{qa}{\omega}}{(\Delta(k a)^2-\omega^2)^2}.
\end{equation}

\section{Third-order susceptibility}
\label{app:chi3}

Here, we work with the Hamiltonian, $H=H_0+H_P(t)$, and briefly outline how to derive an expression of the third-order susceptibility. As before, we write time-independent eigenstates of $H_0$ as $|m\rangle$. 
A time-evolved eigenstate can be derived using perturbation theory
\begin{align}
|\psi_n(t)\rangle = |n\rangle e^{-i \varepsilon_n t}-\sum_{m,\omega}\langle m|H_P(\omega)|n\rangle \frac{e^{-i\varepsilon_n t+i\omega t}}{\varepsilon_m-\varepsilon_n+\omega-i\Gamma/2}|m\rangle+ \nonumber \\
\sum_{k,m,\omega,\omega'}\frac{\langle m|H_P(\omega)|k\rangle \langle k|H_P(\omega')|n\rangle}{(\varepsilon_m-\varepsilon_n+\omega+\omega'-i\Gamma/2)(\varepsilon_k-\varepsilon_n+\omega'-i\Gamma/2)}e^{-i\varepsilon_n t+i\omega t+i\omega' t}|m\rangle -  \nonumber \\
\sum_{k,m,p,\omega,\omega',\omega''}\frac{\langle m|H_P(\omega)|k\rangle \langle k|H_P(\omega')|p\rangle\langle p|H_P(\omega'')|n\rangle e^{-i\varepsilon_n t+i\omega t+i\omega' t+i\omega'' t}}{(\varepsilon_m-\varepsilon_n+\omega+\omega'+\omega''-i\Gamma/2)(\varepsilon_k-\varepsilon_n+\omega'+\omega''-i\Gamma/2)(\varepsilon_p-\varepsilon_n+\omega''-i\Gamma/2)}|m\rangle,
\end{align}
In this appendix, we assume that all transitions are far from the resonance. Therefore, we shall use $\Gamma=0$. As before (for the calculation of the linear susceptibility), we compute the expectation value of the Hamiltonian, $\langle \psi_n(t)| H |\psi_n(t)\rangle$:
\begin{align}
\langle \psi_n(t)| H |\psi_n(t)\rangle = -\sum_{k,m,p,\omega,\omega',\omega'',\omega'''}\frac{ \langle n| H_P(\omega''')|m\rangle\langle m|H_P(\omega)|k\rangle \langle k|H_P(\omega')|p\rangle\langle p|H_P(\omega'')|n\rangle e^{i\omega t+i\omega' t+i\omega'' t+i\omega'''t}}{(\varepsilon_m-\varepsilon_n+\omega+\omega'+\omega''-i\Gamma/2)(\varepsilon_k-\varepsilon_n+\omega'+\omega''-i\Gamma/2)(\varepsilon_p-\varepsilon_n+\omega''-i\Gamma/2)}\nonumber\\
-\sum_{k,m,p,\omega,\omega',\omega'',\omega'''}\frac{ \langle m| H_P(\omega''')|n\rangle\langle m|H_P(\omega)|k\rangle^* \langle k|H_P(\omega')|p\rangle^*\langle p|H_P(\omega'')|n\rangle^* e^{-i\omega t-i\omega' t-i\omega'' t+i\omega'''t}}{(\varepsilon_m-\varepsilon_n+\omega+\omega'+\omega''+i\Gamma/2)(\varepsilon_k-\varepsilon_n+\omega'+\omega''+i\Gamma/2)(\varepsilon_p-\varepsilon_n+\omega''+i\Gamma/2)}\nonumber\\
-\sum_{k,m,p,\omega,\omega',\omega'',\omega'''}\frac{\langle m|H_P(\omega)|n\rangle^*\langle m|H_P(\omega''')|p\rangle\langle p|H_P(\omega'')|k\rangle \langle k|H_P(\omega')|n\rangle e^{-i\omega t+i\omega't+i\omega'' t+i\omega''' t}}{(\varepsilon_p-\varepsilon_n+\omega''+\omega'-i\Gamma/2)(\varepsilon_k-\varepsilon_n+\omega'-i\Gamma/2)(\varepsilon_m-\varepsilon_n+\omega+i\Gamma/2)}\nonumber\\
-\sum_{k,m,p,\omega,\omega',\omega'',\omega'''}\frac{\langle m|H_P(\omega)|k\rangle^* \langle k|H_P(\omega')|n\rangle^*\langle m|H_P(\omega''')|p\rangle\langle p|H_P(\omega'')|n\rangle e^{-i\omega t-i\omega' t+i\omega'' t+i\omega'''t}}{(\varepsilon_m-\varepsilon_n+\omega+\omega'+i\Gamma/2)(\varepsilon_k-\varepsilon_n+\omega'+i\Gamma/2)(\varepsilon_p-\varepsilon_n+\omega''-i\Gamma/2)}.
\end{align}
We average this expression over time to produce an expression suitable for calculating the third-order susceptibility
\begin{align}
\langle \psi_n(t)| H |\psi_n(t)\rangle|_T = -\sum_{k,m,p,\omega,\omega',\omega''}\frac{ \langle n| H_P(-\omega-\omega'-\omega'')|m\rangle\langle m|H_P(\omega)|k\rangle \langle k|H_P(\omega')|p\rangle\langle p|H_P(\omega'')|n\rangle}{(\varepsilon_m-\varepsilon_n+\omega+\omega'+\omega''-i\Gamma/2)(\varepsilon_k-\varepsilon_n+\omega'+\omega''-i\Gamma/2)(\varepsilon_p-\varepsilon_n+\omega''-i\Gamma/2)}\nonumber\\
-\sum_{k,m,p,\omega,\omega',\omega''}\frac{ \langle m| H_P(-\omega-\omega'-\omega'')|n\rangle\langle k|H_P(\omega)|m\rangle \langle p|H_P(\omega')|k\rangle\langle n|H_P(\omega'')|p\rangle}{(\varepsilon_m-\varepsilon_n-\omega-\omega'-\omega''+i\Gamma/2)(\varepsilon_k-\varepsilon_n-\omega'-\omega''+i\Gamma/2)(\varepsilon_p-\varepsilon_n-\omega''+i\Gamma/2)}\nonumber\\
-\sum_{k,m,p,\omega,\omega',\omega''}\frac{\langle n|H_P(\omega)|m\rangle\langle m|H_P(-\omega-\omega'-\omega'')|p\rangle\langle p|H_P(\omega'')|k\rangle \langle k|H_P(\omega')|n\rangle}{(\varepsilon_p-\varepsilon_n+\omega''+\omega'-i\Gamma/2)(\varepsilon_k-\varepsilon_n+\omega'-i\Gamma/2)(\varepsilon_m-\varepsilon_n-\omega+i\Gamma/2)}\nonumber\\
-\sum_{k,m,p,\omega,\omega',\omega''}\frac{\langle k|H_P(\omega)|m\rangle \langle n|H_P(\omega')|k\rangle\langle m|H_P(-\omega-\omega'-\omega'')|p\rangle\langle p|H_P(\omega'')|n\rangle}{(\varepsilon_m-\varepsilon_n-\omega-\omega'+i\Gamma/2)(\varepsilon_k-\varepsilon_n-\omega'+i\Gamma/2)(\varepsilon_p-\varepsilon_n+\omega''-i\Gamma/2)}.
\end{align}
Notice that the sign of the argument is negative only for the total frequency (for $\omega,\omega',\omega''>0$), which leads to the expression presented in the main matter.

Finally, for convenience, we present the general expression for $\bar \chi^3$ that was used in the analysis of THG in the main matter:
\begin{align}
\chi_{ljih}^{(3)}(\Omega,\omega,\omega',\omega'')= A \sum_{\nu,m,n}\frac{d_{g\nu}^h(\omega'')d_{\nu n}^j(\omega)d_{nm}^i(\omega')d_{mg}^l(-\Omega)}{(\varepsilon_{\nu g}-\omega'')(\varepsilon_{ng}+\omega'-\Omega)(\varepsilon_{mg}-\Omega)} +
A \sum_{\nu,m,n}\frac{d_{g\nu}^j(\omega)d_{\nu n}^h(\omega'')d_{nm}^i(\omega')d_{mg}^l(-\Omega)}{(\varepsilon_{\nu g}-\omega)(\varepsilon_{ng}+\omega'-\Omega)(\varepsilon_{mg}-\Omega)} +\nonumber \\
A \sum_{\nu,m,n}\frac{d_{g\nu}^i(\omega')d_{\nu n}^j(\omega)d_{nm}^h(\omega'')d_{mg}^l(-\Omega)}{(\varepsilon_{\nu g}-\omega')(\varepsilon_{ng}+\omega''-\Omega)(\varepsilon_{mg}-\Omega)} +
A \sum_{\nu,m,n}\frac{d_{g\nu}^h(\omega'')d_{\nu n}^i(\omega')d_{nm}^j(\omega)d_{mg}^l(-\Omega)}{(\varepsilon_{\nu g}-\omega'')(\varepsilon_{ng}+\omega-\Omega)(\varepsilon_{mg}-\Omega)}+\nonumber\\
A \sum_{\nu,m,n}\frac{d_{g\nu}^j(\omega)d_{\nu n}^i(\omega')d_{nm}^h(\omega'')d_{mg}^l(-\Omega)}{(\varepsilon_{\nu g}-\omega)(\varepsilon_{ng}+\omega''-\Omega)(\varepsilon_{mg}-\Omega)}+A \sum_{\nu,m,n}\frac{d_{g\nu}^i(\omega')d_{\nu n}^h(\omega'')d_{nm}^j(\omega)d_{mg}^l(-\Omega)}{(\varepsilon_{\nu g}-\omega')(\varepsilon_{ng}+\omega-\Omega)(\varepsilon_{mg}-\Omega)}.
\end{align}
The corresponding expression for $\bar \chi^{(3)}$ reads as follows
\begin{align}
\bar \chi_{ljih}^{(3)}(\Omega,\omega,\omega',\omega'')\simeq  A\int\frac{\mathrm{d}\mathbf{k}}{(2\pi)^3} \frac{f_l(-\Omega)f_j(\omega)f_i(-\omega')f_h(-\omega'')\langle\sigma_h\sigma_j\sigma_i\sigma_l\rangle_{\Uparrow}}{(\Delta(ka)-\omega'')(\omega'-\Omega)(\Delta(ka)-\Omega)}\nonumber\\+A\int\frac{\mathrm{d}\mathbf{k}}{(2\pi)^3} \frac{f_l(-\Omega)f_j(-\omega)f_i(\omega')f_h(-\omega'')\langle \sigma_h\sigma_i\sigma_j\sigma_l \rangle_{\Uparrow}}{(\Delta(ka)-\omega'')(\omega-\Omega)(\Delta(ka)-\Omega)}\nonumber+A\int\frac{\mathrm{d}\mathbf{k}}{(2\pi)^3} \frac{f_l(-\Omega)f_j(\omega)f_i(-\omega')f_h(-\omega'')\langle \sigma_i\sigma_j\sigma_h\sigma_l\rangle_{\Uparrow}}{(\Delta(ka)-\omega')(\omega''-\Omega)(\Delta(ka)-\Omega)}\nonumber\\+A\int\frac{\mathrm{d}\mathbf{k}}{(2\pi)^3} \frac{f_l(-\Omega)f_j(-\omega)f_i(-\omega')f_h(\omega'')\langle \sigma_j\sigma_h\sigma_i\sigma_l\rangle_{\Uparrow}}{(\Delta(ka)-\omega)(\omega'-\Omega)(\Delta(ka)-\Omega)}\nonumber+A\int\frac{\mathrm{d}\mathbf{k}}{(2\pi)^3} \frac{f_l(-\Omega)f_j(-\omega)f_i(\omega')f_h(-\omega'')\langle \sigma_j\sigma_i\sigma_h\sigma_l \rangle_{\Uparrow}}{(\Delta(ka)-\omega)(\omega''-\Omega)(\Delta(ka)-\Omega)}\nonumber\\+A\int\frac{\mathrm{d}\mathbf{k}}{(2\pi)^3} \frac{f_l(-\Omega)f_j(-\omega)f_i(-\omega')f_h(\omega'')\langle \sigma_i\sigma_h\sigma_j\sigma_l\rangle_{\Uparrow}}{(\Delta(ka)-\omega')(\omega-\Omega)(\Delta(ka)-\Omega)}.
\end{align}

\end{widetext}

\bibliography{refs_method}

\end{document}